\documentclass[12pt]{article}
\begin{document}
%The following macro is from world_sci.sty, originally written for DPF91

\catcode`@=11
% Collapse citation numbers to ranges.  Non-numeric and undefined labels
% are handled.  No sorting is done.  E.g., 1,3,2,3,4,5,foo,1,2,3,?,4,5
% gives 1,3,2-5,foo,1-3,?,4,5
\newcount\@tempcntc
\def\@citex[#1]#2{\if@filesw\immediate\write\@auxout{\string\citation{#2}}\fi
  \@tempcnta\z@\@tempcntb\m@ne\def\@citea{}\@cite{\@for\@citeb:=#2\do
    {\@ifundefined
       {b@\@citeb}{\@citeo\@tempcntb\m@ne\@citea\def\@citea{,}{\bf ?}\@warning
       {Citation `\@citeb' on page \thepage \space undefined}}%
    {\setbox\z@\hbox{\global\@tempcntc0\csname b@\@citeb\endcsname\relax}%
     \ifnum\@tempcntc=\z@ \@citeo\@tempcntb\m@ne
       \@citea\def\@citea{,}\hbox{\csname b@\@citeb\endcsname}%
     \else
      \advance\@tempcntb\@ne
      \ifnum\@tempcntb=\@tempcntc
      \else\advance\@tempcntb\m@ne\@citeo
      \@tempcnta\@tempcntc\@tempcntb\@tempcntc\fi\fi}}\@citeo}{#1}}
\def\@citeo{\ifnum\@tempcnta>\@tempcntb\else\@citea\def\@citea{,}%
  \ifnum\@tempcnta=\@tempcntb\the\@tempcnta\else
   {\advance\@tempcnta\@ne\ifnum\@tempcnta=\@tempcntb \else \def\@citea{--}\fi
    \advance\@tempcnta\m@ne\the\@tempcnta\@citea\the\@tempcntb}\fi\fi}
\catcode`@=12
% 
%Preprint(complete version)
\linespread{1.3}
\title{A solution of the measurement problem in quantum mechanics
by using a variable hidden in Newtonian mechanics\footnote{revised
version}}
\author{{\sc Jae-Hyung Myung}\\
II.Institut f\"ur Theoretische Physik, Universit\"at Hamburg,\\
D-22761 Hamburg , Germany}
\date{22/03/1996}
\maketitle
\begin{abstract}
The problem of a correct description of the physical phenomena of the Heisenberg uncertainty relation 
is solved by using a variable hidden in Newtonian mechanics.\\
\end{abstract}
\vspace{5cm}
%%%UH-ITHPII-95-1
\newpage
\section{Introduction}
The aim of this work is to solve the measurement problem in quantum
mechanics. Since the problem was introduced by Einstein et al.(EPR)(see Ref.~\cite{Einstein:1935we}),
although many efforts have been devoted to the solution of the
problem, to date no conclusive solution has been found.\\
In quantum mechanics two physical quantities described by non-commuting
operators cannot be simultaneously measured with perfect accuracy, and
the relation between these two quantities can be derived from a wave
function. The problem put forward by EPR is that the description of
physical reality as given by the wave function is not complete. The most
important part in the EPR paper is the criteron of reality: "If, without
in any way disturbing a system, we can predict with certainty (i.e.,
with probability equal to unity) the value of a physical quantity, then
there exists an element of physical reality corresponding to this
physical quantity." According to the EPR arguments quantum mechanics is
not a complete theory, since there is not the reality element. This
criterion implies the existence of an additional (hidden)
variable.\\
Another important point in the paper is the problem of locality: "Since
at the time of measurement the two systems no longer interact, no real
change can take place in the second system in consequence of anything
that may be done to the first system." The EPR arguments of reality and
locality were developed into the local hidden variable theory in quantum
mechanics (see Refs.~\cite{Bohm:1952hl,Bohm:1952sk,Bell:1965rt,Bell:1966fg,Bohm:1966er,Belinfante:1973hj}). 
According to the authors, quantum mechanics needs a
variable for a correct description at small distance.\\
The Bell theorem deduced from the locality assumptions was generalized in Refs.~\cite{Clauser:1969kl,
Clauser:1974sm,Garuccio:1981sd} and experimentally tested in Refs.~\cite{Freedman:1972af,Clauser:1976cv,Clauser:1976bn,
Aspect:1981cs,Perrie:1985gh}. The weak inequalities are
fully compartible with the experimental results, while the strong
inequalitites are violated (see Ref.~\cite{Home:1991fg}). The EPR arguments concerning the
correct description of physical reality of the wave function are
not reasonable, since the physical theories described by a wave
function are in good agreement with experiments.\\
EPR have not directly referred to the uncertainty relation in their
paper. However, since one of the authors, Einstein, has been against
Bohr's interpretation of the relation, in this paper we confine the
measurement problem to the uncertainty relation. That is, we solve the
problem of a correct description of the physical phenomena of the
relation by making use of a variable hidden in Newtonian mechanics. As
we begin to see in next section, the hidden variable plays a crucial
role in the solution of the problem. In this case the hidden variable
is a physical quantity that is omited from the object of measurement,
since the quantity is hidden in other quantities. That is, the variable
in question is a variable hidden in Newtonian mechanics.\\
A complete physical theory is in a precise (one-one) correspondence with
the object that is being described. However, a hidden variable makes a
physical theory correspond to two objects. The uncertainty relation
describes the relation between momentum and position of a particle. If
an unknown variable is hidden in one of these two quantities, we need
another physical theory to describe the relation between the hidden
variable and the complementary variable.\\
The uncertainty relation is a measurement theory of two quantities of
a particle, and a kernel in the measurement theory is interpretation.
Whether we can simultaneously know two physical quantities with perfect
accuracy or not, depends only on the interpretation. Various measurement
theories were suggested by many authors (Ref.~\cite{Wheeler:1983sg}). The most exact of these
theories is the principle of complementarity of Bohr (Ref.~\cite{Bohr:1935dl}) and his
interpretation. If we arrange matters so that a quantity in the
relation is small, another will be large. Thus, we can simultaneously
measure these two quantities with limited accuracy. If any two
quantities are proportional, the measurement processes are entirely
different from the former situation. If we arrange matters so that a
quantity is small, another will be also small. Thus, we can
simultaneously measure these two quantities with perfect accuracy. The
variable hidden in classical mechanics is proportional to another known
variable. In this work we find the hidden variable and solve the
measurement problem by using this variable. \\
This paper is organized as follows. In section 2 it is shown that a
variable hidden in Newtonian mechanics plays a key role in the solution
of the problem, and in section 3 the hidden variable (the quantity of
motion) is redefined. In section 4 we solve the measurement problem in
quantum mechanics by using the new variable. Section 5 is devoted to conclusions.
\section{The uncertainty relation and a variable hidden in Newtonian mechanics}
The uncertainty relation for a bound electron in an atom is given as
\begin{equation}
\triangle{P}\triangle{x} \sim nh,
\end{equation}
if n means the quantum number of the stationary state (Ref.~\cite{Heisenberg:1958dk}).
The relation (1) can be rewritten as follows
\begin{equation}
\triangle{P}\triangle{x} \geq h.
\end{equation}
If we give up the knowledge of the stationary state, that is, if the
electron is practically regarded as free, the uncertainty relation
is given as
\begin{equation}
\triangle{P}\triangle{x} \sim h.
\end{equation}
The relations (1)-(3) say: If we arrange matters so that $\triangle{x}$
is small, $\triangle{P}$ will be large. If we reduce $\triangle{P}$ in
some way, $\triangle{x}$ will be large. Therefore, we cannot
simultaneously measure both momentum and position with perfect accuracy.
We consider the uncertainty relation (3) in this work. The relation was
confirmed by experiment, for example, in the collision of particles in
an accelerator. Therefore, we are convinced that the relation completely
and formally describes the physical phenomena (physical reality) of a
moving particle.\\
To solve the problem of a correct description of the physical phenomena
of the relation by an approach of the hidden variable theoy described in
section 1, we must consider the relation and the quantities in the
relation conceptually. That is, we must find the variable hidden in
Newtonian mechanics. In the relation, the concept of P is not clear.  P
(momentum) has two meanings, the quantity of motion and the force of a
moving body. Let us consider these two concepts separately.\\
If we consider P as the force of a moving particle, then the relation
describes the physical phenomena in the conceptional sense correctly,
that is, the relation between force and position of a particle can be
described through the uncertainty relation (3). On the other hand, if
we consider P as the quantity of motion of a particle, then the
relation does not describe the phenomena correctly, that is, the
relation between the quantity of motion and position of particle cannot
be described through the uncertainty relation, since  the quantity of
motion is proportional to the position. Therefore, the variable (the
quantity of motion) hidden in Newtonian mechanics plays a key role for
the solution of the problem of a correct description of the physical
phenomena of the uncertainty relation. We must define the concept of
the quantity of motion to solve the problem exactly. This is done in
next section.
\section{New foundations of classical mechanics}
We define the product of mass and velocity as momentum. The term
momentum has two meanings, the quantity of motion ('quantitus motus')
and impulse ('impetus'). The former originated from Descartes (Ref~\cite{Descartes:1982ak}) and
Newton (Ref.~\cite{Newton:1964fl}) and the latter from Leibnitz (Ref.~\cite{Leibnitz:1982dm}). 
However, these two concepts represent different physical quantities.\\
The measure of the magnitude of the impulse (or force) of a moving body
is magnitude of velocity, while the measure of the quantity of motion
is the path length covered by the body. That is, the quantity of motion
is proportional not to the magnitude of velocity but to the covered
path length.\\
Motion is the variation of position in space. Therefore, the quantity
of motion is proportional to the quantity of the variation of position,
which can be exactly expressed through the covered path length. Thus,
the quantity of motion Q is defined as the product of mass m and the
covered path length l.\\
The quantity of motion is a scalar quantity, since the covered path
length is a scalar. The quantity of motion Q of a body is described
through an integral formula. If a body of mass m moves along a path
from position a to position b in the x-y coordinate system and dQ (=m
dl) is the quantity of infinitesimal motion, then the quantity of
motion of the body is given by
\begin{equation}
Q = \int dQ = ml.
\end{equation}
The product of mass and velocity is a formal description of a moving
body which exerts a force on another body during a collision. Therefore,
the product is defined as the force or the momentum  of a moving body.
The magnitude of the momentum P of a moving body is proportional to the
rate of change of the quantity of the motion Q:
\begin{equation}
P = \frac{dQ}{dt}.
\end{equation}
The product of mass and acceleration is a formal description of an
accelerating body which exerts a force on another body during a
collision. Therefore, the product is defined as the force of an
accelerating body. The further work in classical mechanics is done
in the appendix.
\section{A solution of the measurement problem in quantum mechanics}
In this section we solve the problem of a correct description of
the physical phenomena of the uncertainty relation by using the new
variable Q. As was referred in section 2, the uncertainty relation
correctly describes the physical phenomena formally. However, the
relation is not complete in the conceptional sense because of the hidden
variable Q. We need another theory describing the relation between
the quantity of motion Q and position x.\\
Let us consider the relation between two physical quantities
$\triangle{Q}$ and $\triangle{x}$ of a particle according to the new
definition of the quantity of motion. If a particle moves in the
direction of x in the x-y coordinate system, then we know from the
motion of the wave form of the particle that the position
$\triangle{x}$ is proportional to the path length $\triangle{l}
$ covered by the particle (see Fig.1) and from the definition of Q,
that $\triangle{l}$ is proportional to $\triangle{Q}$. Consequently,
these two proportional relations mean that $\triangle {Q}$ is
proportional to $\triangle{x}$. That is, from the two relations
$\triangle{l} = n \triangle{x}$ and $ \triangle{Q} = m \triangle
{l}$, the relation
\begin{equation}
\triangle{Q} = k \triangle{x} (k=mn)
\end{equation}
results. The proportional relation between $\triangle{Q}$ and $\triangle
{x}$ means that $\triangle{Q}$ differs from $\triangle{x}$ by a
proportional constant k. This fact says: If we arrange matters so that
$\triangle{x}$ is small, $\triangle{Q}$ will be also
small. If we reduce $\triangle{Q}$, $\triangle{x}$ will also be reduced.
Therefore, if we know the position of a particle, then we can
also know accurately the quantity of motion of the particle and
conversely, if we know the quantity of motion of a particle, then we
can also know accurately the position of the particle. In other words,
we can measure simultaneously two physical quantities, i.e., the
quantity of motion and the position of a particle with perfect
accuracy. Hereafter we call the relation (6) the certainty
relation.\\
The relation between two observables Q and x can be described through a
commutation relation. We regard the observable Q as a trivial operator.
Two trivial operators Q and x then commute:
\begin{equation}
[Q, x] = 0.
\end{equation}
The uncertainty relation between $\triangle{P}$ and $\triangle{x}$
is a well established theory in the formal sense. In this work
we deduce the relation from the de Broglie relation to compare with
relation (6) or (7). Let us consider the de Broglie relation $P\lambda
= h$. If we take the uncertainty $\triangle{P} (\sim P)$ in momentum P,
the relation becomes
\begin{equation}
\triangle{P}\lambda \sim h.
\end{equation}
If we choose one wavelength as the uncertainty $\triangle{x}$ in
position (locality condition), we can insert $\triangle{x}$ instead of
$\lambda$ in the relation and obtain the uncertainty relation
\begin{equation}
\triangle{P}\triangle{x} \sim h.
\end{equation}
The de Broglie relation thus implies the uncertainty relation between
momentum and position of a  particle. If we arrange matters so that
$\triangle {P}$ is small, $\triangle{x}(or \lambda)$ will be large.
If we reduce $\triangle{x}(or \lambda)$ in some way, $\triangle{P}$
will be large. Therefore, we cannot simultaneously measure monentum
and position (or wavelength) of a particle with perfect accuracy. The
relation between two observables P and x is also described through the
commutation relation. The operator P does not commute with x:
\begin{equation}
[P, x] =i\hbar,
\end{equation}
where the operator P is given by
\begin{equation}
 P =i\hbar\frac{\partial}{\partial x}.
\end{equation}
The relation (9) or (10) is entirely different from the relation (6) or
(7). Therefore, the problem of a correct description of the physical
phenomena of the uncertainty relation is due not to the incompleteness
of quantum mechanics but to the variable Q hidden in Newtonian
mechanics. That is, although the uncertainty relation correctly
describes the physical phenomena of a moving particle formally, the
relation is not complete in the conceptional sense because of the
hidden variable Q. The relation between the new variable Q and position
x is given by the relations (6) and (7).
\section{Conclusions}
In the present work we confined the measurement problem to the problem
of a correct description of the physical phenomena of the uncertainty
relation and solved the problem. The uncertainty relation is complete in
the formal sense. However, if we consider the relation conceptually,
the relation is not complete because of the variable Q hidden in
Newtonian mechanics. The quantity of motion Q and momentum P are
different physical quantities. The relation between the quantity of
motion and position of a particle is described through the certainty
relation, while the relation between momentum and position of the
particle is described through the uncertainty relation. From the
locality condition we see that quantum mechanics is a local
theory.\\
Whether we can simultaneously measure two physical quantities with
perfect accuracy or not, depends only on the interpretation. If we
arrange matters so that a quantity in the certainty relation is small,
another will be also small. Therefore, we can simultaneously measure
two quantities in the relation with perfect accuracy. On the other
hand, if we reduce a quantity in the uncertainty relation in some way,
another will be large. Therefore, we cannot simultaneously measure two
quantities in the relation with perfect accuracy.
\vspace{5mm}
Acknowledgement\\
I will not submit this paper to any physical journal because
of the prejudice of editors.
\begin{appendix}
\section {Three laws of motion in classical mechanics}
According to the new definition of the quantity of motion, momentum and
force, we formulate three laws of motion in classical mechanics.
The magnitude of the momentum P of a moving body is proportional to the
rate of change of the quantity of the motion Q:
\begin{equation}
P = \frac{dQ}{dt}.
\end{equation}
This relation is called the first law of motion. The relation between
momentum and force is given as follows. The magnitude of the force F of
an accelerating body is proportional to the rate of change of P:
\begin{equation}
F = \frac{dP}{dt}.
\end{equation}
The above relation is the second law of motion. This law is different
from the Newton's second law, since F in new mechanics is no external
force. From the above both relations  we get the equation of motion
\begin{equation}
F = \frac{d^2Q}{dt^2}.
\end{equation}
When a body is accelerated by an external field, the magnitude of the
force of the accelerating body is equal to that of the force that the
field exerts on the body. This is the third law of motion. The law
gives the equation of motion for a body moving in an external field.
This new mechanics can be applied not only to microscopic systems,
but also to macroscopic systems.
\end{appendix}

\newpage
%\listoffigures
Figure caption\\
Fig.1 Wave motion of a particle. The covered path length $\triangle{l}$
is proportional to the position $\triangle{x}$, and the wavelength
$\lambda$ is equal to the position $\triangle{x}$. 
\end{document}